# Octave-Tunable Magnetostatic Wave YIG Resonators on a Chip

Sen Dai, Sunil A. Bhave, *Senior Member, IEEE*, and Renyuan Wang, *Member, IEEE*

*Abstract*—**We have designed, fabricated, and characterized magnetostatic wave (MSW) resonators on a chip. The resonators are fabricated by patterning single-crystal yttrium iron garnet (YIG) film on a gadolinium gallium garnet (GGG) substrate and excited by loop-inductor transducers. We achieved this technology breakthrough by developing a YIG film etching process and fabricating thick aluminum coplanar waveguide (CPW) inductor loop around each resonator to individually address and excite MSWs. At 4.77 GHz, the 0.68-mm$^2$ resonator achieves a quality factor ($Q$) > 5000 with a bias field of 987 Oe. We also demonstrate YIG resonator tuning by more than one octave from 3.63 to 7.63 GHz by applying an in-plane external magnetic field. The measured quality factor of the resonator is consistently over 3000 above 4 GHz. The micromachining technology enables the fabrication of multiple single- and two-port YIG resonators on the same chip with all resonators demonstrating octave tunability and high $Q$.**

## I. Introduction

THE advent of 5G and the desire for large bandwidth has brought the 3–30-GHz band into prominence [1]. RF MEMS piezoelectric film bulk acoustic resonators (FBARs) [2], the gold standard of 4G filter technology, do not scale favorably with 5G RF communication [3] because of reduced thickness, high metal resistance, and challenging lithography. On the other hand, electromagnetic (EM) wave-based resonators, such as microstrip lines, 3-D micromachined coaxial lines, and evanescent cavities [4], [5], are too large for chip-scale integration.

Magnetostatic wave (MSW) resonators and filters are a promising technology to fill this gap [6]. MSWs exist in ferromagnetic/ferrimagnetic materials. It is a lattice wave where the lattice consists of electron spin precessions. These waves possess two salient features making them attractive for realizing chip-scale resonators and filters in the super-high-frequency (SHF) band. First, the group velocity is on the order of 1000 km/s and is a strong function of magnetic bias applied to the material [7]. Therefore, the device size does not scale to extremely small dimensions with increased operating frequency. Second, the material loss limited $Q$ of MSW resonator is theoretically frequency independent. Single-crystal yttrium iron garnet (YIG) exhibits the lowest damping for MSW, with a material loss limited $Q > 10\,000$ for frequencies in the UHF to *Ka* bands [8], [9] has demonstrated a YIG MSW resonator can reach $Q > 3000$ in the *X*-band.

State-of-the-art MSW resonators are constructed from bonding a YIG-on-gadolinium gallium garnet (GGG) substrate on to another low dielectric loss substrate with $\lambda/4$ or $\lambda/2$ planar transmission lines to excite the MSW resonant mode [6], [9]. Such an approach leads to a centimeter-scale resonator forfeiting the YIG's advantage over a conventional EM resonator. This poses challenges in monolithic integration and miniaturization of multiple MSW devices and limits its application in higher order MSW filters and multiplexers. In this article, we designed a novel MSW resonator that consists of a YIG thin-film mesa as well as a new loop-inductor transducer structure to efficiently excite the MSW [see Fig. 1(a)], leading to significantly reduced resonator size. We also developed a new microfabrication process to fabricate multiple MSW resonators on the same chip, as shown in Fig. 1(b). The combination of novel microfabrication process with the significantly reduced resonator size provides the freedom to design different MSW resonators with different wavelengths on a single chip so that they have different resonant frequencies even with the same magnetic bias. This provides the potential of monolithic high-order MSW filters and multiplexers with a small external magnet.

## II. Resonator Design and Modeling

In this work, 3-$\mu$m liquid-phase epitaxy (LPE) YIG film grown on 500-$\mu$m (111)-oriented GGG substrate was used. The saturation magnetization ($M_s$) of YIG is ~1750 G, and the gyromagnetic ratio $\gamma \mu_0$ is 2.8 MHz/G.

A thin-film ferromagnetic/ferrimagnetic structure can support three different types of MSW—magnetostatic forward volume wave (MFVW), magnetostatic backward volume wave (MBVW), and magnetostatic surface wave (MSSW) [7]. With the in-plane external magnetic bias field applied in parallel

Manuscript received January 27, 2020; accepted June 1, 2020. Date of publication June 4, 2020; date of current version October 26, 2020. This work was supported by Defense Advanced Research Projects Agency (DARPA) under Contract HR0011-19-C-0017. Purdue co-authors also acknowledge support under Semiconductor Research Corporation (SRC) Contract 2018-LM-2830. The views, opinions and/or findings expressed are those of the author and should not be interpreted as representing the official views or policies of the Department of Defense or the U.S. Government. This manuscript was approved for public release; distribution statement A; distribution unlimited. This manuscript is not export controlled per ES-FL-011720-0016. *(Corresponding author: Renyuan Wang.)*
Sen Dai is with the Department of Physics and Astronomy, Purdue University, West Lafayette, IN 47907 USA.
Sunil A. Bhave is with the School of Electrical and Computer Engineering, Purdue University, West Lafayette, IN 47907 USA.
Renyuan Wang is with FAST Labs, BAE Systems, Inc., Nashua, NH 03060 USA (e-mail: renyuan.wang@baesystems.com).

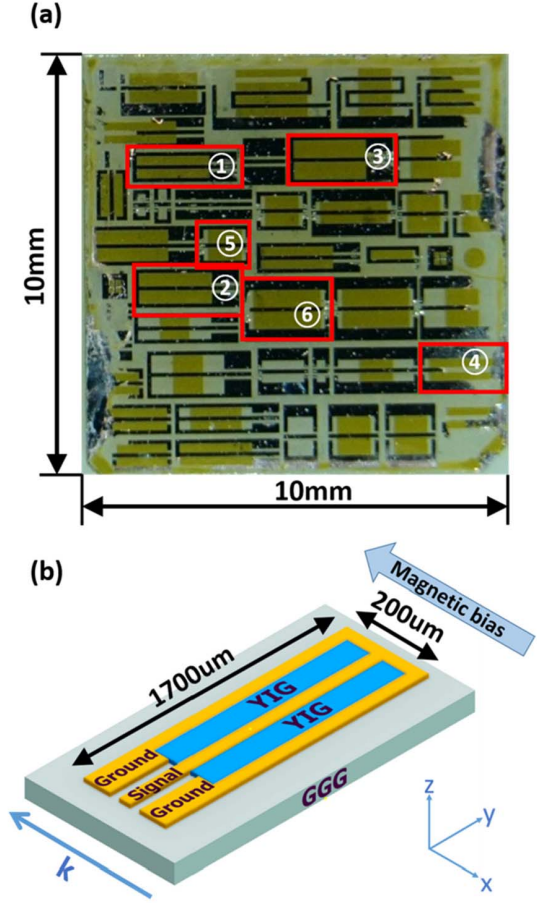

Fig. 1. (a) Top view of the chip with multiple MSW YIG resonators that are designed to operate from 4.54 to 4.58 GHz with the same 900-Oe dc magnetic bias [Devices ①–⑥]. (b) Schematic of MSW resonator ① as marked in red.

In contrast with conventional method of exciting MBVW through centimeter long transmission lines, a simple planar loop-inductor design is proposed. The inductor loop generates an RF magnetic field perpendicular to the YIG film. As the excitation efficiency of MSW depends on the overlap integral between the RF $H$-field and the MSW mode profile, the inductor loop provides strong coupling as well as realizes a much smaller resonator size.

A simple analogy between bulk acoustic wave and MBVW could be used to further illustrate the working principle of MBVW resonator and the function of inductor loop as follows.

1) An external magnetic bias $H$ above the saturation field is needed to align the magnetic spin momentum ($M$) with the external magnetic bias, which corresponds to the poling process of piezomaterial using an electric field $E$ to reach saturation polarization ($P$) regime.
2) An open circuit, as a capacitor, is used to excite the acoustic wave and the driving force is the charge-induced electric field $E$ perturbation on the polarization $P$. In comparison, a short circuit, as an inductor loop we proposed here, is used to excite the spin wave and the driving force is the RF current-induced magnetic field $H$ perturbation on the magnetic momentum $M$.
3) In piezomaterial, this electric field ($E$)-induced perturbation on $P$ induces a coherent movement of atoms of the lattice out of their equilibrium positions, which is called an acoustic wave. In ferromagnetic/ferrimagnetic material, the magnetic field ($h$)-induced perturbation on $M$ induces a coherent spin precession, which is called MSW.

Despite the similarities between acoustic wave and MSW, MSW resonator possesses a few advantages over acoustic wave resonators in the SHF band. First, typical magnetic wave velocity is around $1 \times 10^5$ to $1 \times 10^6$ m/s, leading to a device size of a few millimeters to submillimeters and is easily tuned by an external magnetic bias field to fit different sizes and different frequencies. However, a typical acoustic wave velocity is around 5000–10 000 m/s, requiring a submicrometer-level fabrication and is hard to tune because it requires changing Young's modulus of the piezomaterial. Second, in many types of acoustic resonators, the material loss limited resonator $f \times Q$ product is generally a constant, resulting in intrinsic quality factor degradation at higher frequency [12]. For ferrimagnetic/ferromagnetic material, the material loss limited $Q$ of MSW resonator is theoretically frequency independent, as no thermoelastic damping exists in coherent spin precession. Finally, while the acoustic wave exists in all solid materials, MSW only exists in ferromagnetic/ferrimagnetic materials. Therefore, the material and design of the transducer for acoustic wave devices need to be carefully optimized to achieve a balance between material acoustic loss, energy confinement, and electrical resistance. On the other hand, for MSW, many low resistivity metal materials do not support MSW transportation. This significantly relaxes the constraint for optimizing toward low parasitic electrical resistance from the transducers.

The designed MSW resonator devices consist of two YIG film mesas wrapped around by electrode loop inductors with the wavenumber $k$, MBVW will be excited in the YIG film. Suppose that the YIG film has an infinite lateral size, thus ignoring the in-plane boundary, the wave amplitude distributes sinusoidally through the volume of the film, and the dispersion relation of the lowest order mode ($n = 1$) is

$$\omega^2 = \omega_0 \left[ \omega_0 + \omega_M \left( \frac{1 - e^{-kd}}{kd} \right) \right] \quad (1)$$

where $\omega_0 = \mu_0 \gamma H_{\text{eff}}^{\text{dc}}$, $\omega_M = \mu_0 \gamma M_S$, $\omega$ is the frequency of the MBVW, $k$ is the wave vector of the MBVW and has an in-plane direction, $d$ is the film thickness, $\gamma$ is the gyromagnetic ratio that is fundamentally related to electron charge to mass ratio, $\mu_0$ is the vacuum permeability, $H_{\text{eff}}^{\text{dc}}$ is the magnitude of the effective torque-exerting dc bias field internal to the material, and $M_S$ is the magnitude of the saturation magnetization [10]. For the lowest order mode considered in this manuscript, $kd \ll 1$. Therefore, for a fixed wavelength, the frequency tuning sensitivity is

$$\frac{d\omega}{dH_{\text{eff}}^{\text{dc}}} = \frac{d\omega}{d\omega_0} \frac{d\omega_0}{dH_{\text{eff}}^{\text{dc}}} \approx \frac{1}{2} \frac{2\omega_0 + \omega_M}{(\omega_0^2 + \omega_0 \omega_M)^{\frac{1}{2}}} \mu_0 \gamma. \quad (2)$$

The saturation magnetization of YIG film is ∼1750 G [11], and the gyromagnetic ratio $\gamma \mu_0$ is 2.8 MHz/G.

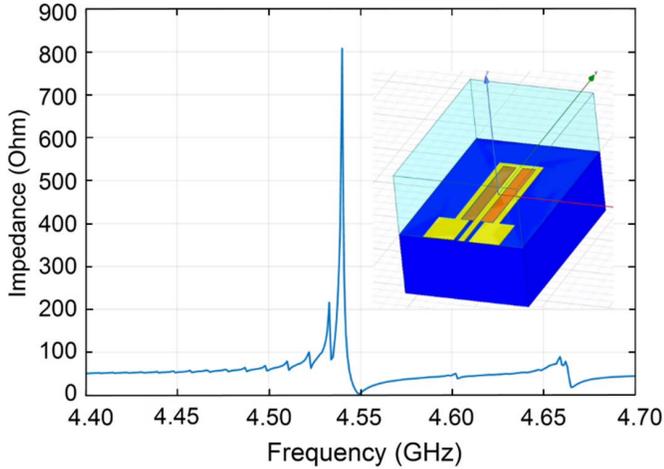

Fig. 2. Device performance simulation using HFSS with the magnetic bias of 950 Oe. The simulated structure mimics the exact physical construct of actual MBVW resonator where two YIG islands sit on a GGG substrate with loop inductors wrapped around them. Inset: schematic of the simulated structure.

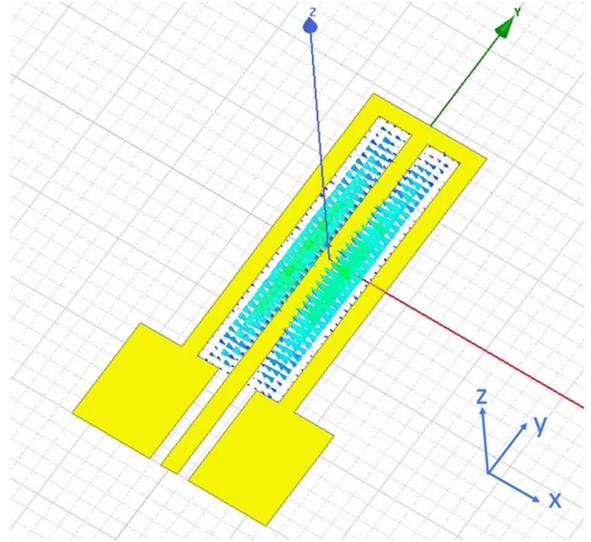

Fig. 3. Mode profile (time-varying component of the magnetization vector) of the MSW in the proposed structure.

(see Fig. 2). As will be described in Section III, the YIG mesas are formed by a unique ion mill etching technique. Because the GGG substrate does not support spin-wave transportation, this provides 3-D energy confinement of the MSW energy (Fig. 3) similar to that of AlN FBARs. A single YIG film island and its associated loop inductor can be considered as a single MSW resonator, and the device in Fig. 2 can be considered as two MSW resonators connected in parallel. Such a balanced structure is to facilitate the characterization of these device, to avoid the parasitics from converting the balanced RF signal from the coplanar waveguide (CPW) to a single-ended resonator structure. The width of the YIG resonator is 200 $\mu$m, and the length of the YIG resonator is 1700 $\mu$m. The width of the CPW trace is 100 $\mu$m. As submicrometer-thick epitaxial YIG films tend to exhibit higher intrinsic damping for MSW due to the crystal defects [13], we opt for a 3-$\mu$m-thick YIG film to optimize toward high $Q$ operation. In addition, the transducer is intentionally separated from the YIG mesa by 5 $\mu$m to prevent potential MSW damping caused by spin-wave pumping from YIG to metallic electrode material [14].

## III. Fabrication

The key process steps for the MSW resonator fabrication are shown in Fig. 4. A layer of photoresist was patterned on the YIG film as mask for ion mill etching. As the etching selectivity between photoresist and YIG was tested to be close to 1:1, the photoresist thickness was chosen to be 5 $\mu$m in order to ensure sufficient masking as well as prevent the transfer of PR surface morphology to the YIG etching sidewall. With an optimized recipe, we were able to achieve an etching rate of 32.6 nm/min and at the same time retain a vertical sidewall angle with intermissive cooling cycles to avoid PR burning. After ion milling, the chip was transferred to acetone and sonicated for 30 min to remove the hardened PR mask. The hardening of PR was due to Ar plasma exposure and high-temperature during etching. Another 30 min of phosphoric acid soak at 80 °C was implemented to remove most of the resputtered YIG around the sidewall. The SEM of the etched YIG sidewall is shown in Fig. 5. A 2-$\mu$m-thick Al electrode was defined by the liftoff process to minimize electrical resistive loss. To ensure the proper lift-off of the 2-$\mu$m-thick metal, we developed an ultrathick PR recipe. The Al electrode is deposited using e-beam evaporation with 10-min cooling cycles between each of the 400-nm deposition intervals to avoid PR overheating and repeated metal soak process to facilitate source metal reflow. Fig. 5 shows the SEM of the fabricated MW resonator device as well as the top view of one resonator device.

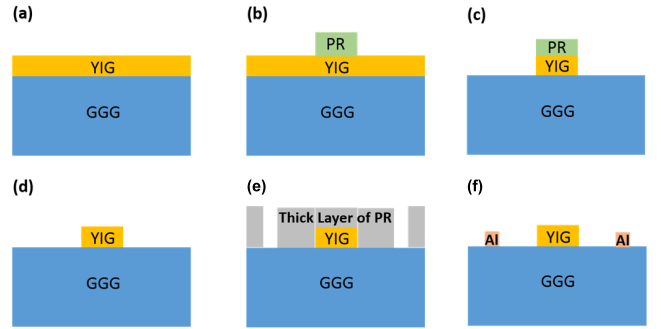

Fig. 4. (a) 3-$\mu$m single-crystal (111) YIG is grown on the GGG substrate. (b) Photoresist with 5-$\mu$m thickness has been patterned as a mask for YIG ion milling. (c) YIG is etched by ion milling at an etching rate of 32.6 nm/min with cooling cycles. (d) PR mask and resputtered YIG is removed for clean YIG patterning. (e) Thick photoresist mask has been patterned through lithography. (f) 2 $\mu$m of Al is e-beam deposited followed by a liftoff process.

## IV. Measurement

A three-axis projection magnet (GMW Magnet System 5201 Model) is used to generate the magnetic bias field and the device is placed 2 mm above the center of the magnet projector to ensure that only in-plane magnetic field is applied to the device. The magnetic field is calibrated using a three-axis Hall sensor. The measured magnetic field uniformity is within $\pm 1\%$ over an area of 20 mm$^2$, where the device is placed to ensure that it is uniformly biased. The scattering parameters of the devices were measured using a network analyzer (Agilent PNA-L N5230A) with 20 000 sampling points in a 2-GHz scan with a resolution bandwidth of 100 kHz (see Fig. 6).

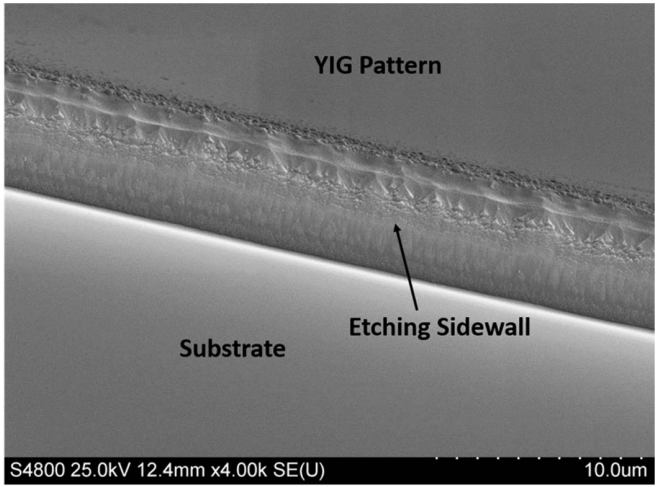

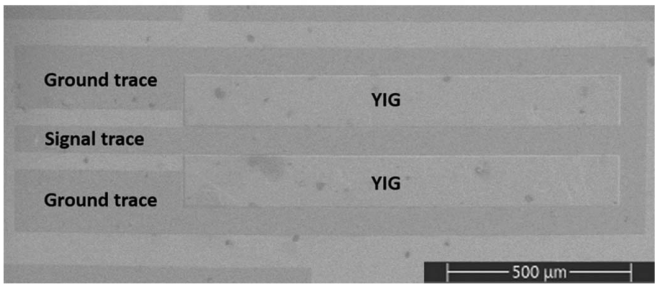

Fig. 5. SEM images of YIG ion milling etching sidewall view (top) and top view of the MSW resonator (bottom).

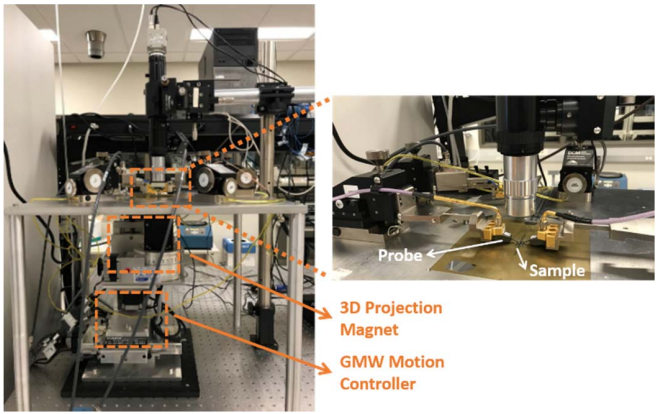

Fig. 6. Photograph of the MSW resonator testing setup (left) and zoomed-in view of RF probe station (right).

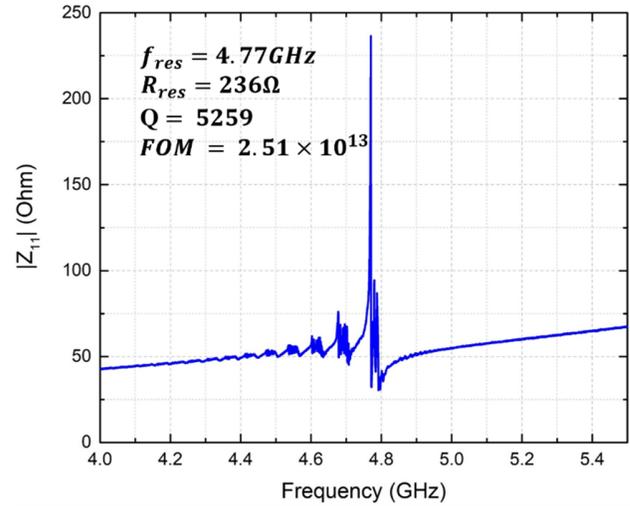

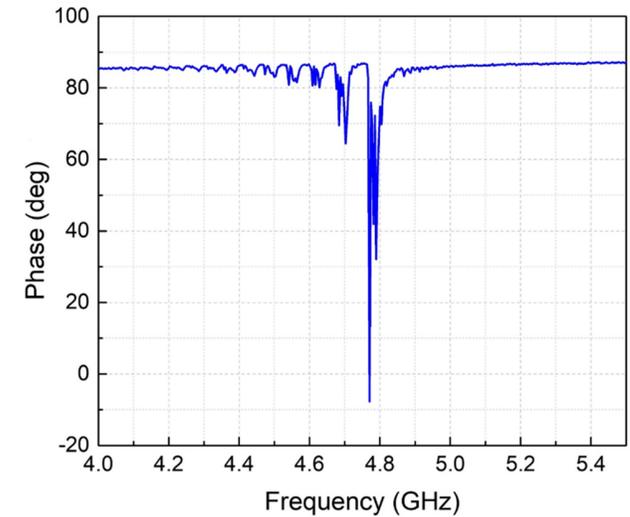

Fig. 7. Magnitude (top) and phase (bottom) of the impedance of the one-port MSW resonator.

The impedance (as measured, no deembedding was performed) of a one-port resonator from 4.0 to 5.5 GHz with a magnetic bias of 987 Oe is measured and shown in Fig. 7. The measured response matches with our simulations. The main resonant frequency is at 4.770 GHz and the 3-dB bandwidth is 0.907 MHz, resulting in a quality factor $Q = f_{\text{resonance}}/\Delta f_{3\,\text{dB}}$ of 5259. To the best of our knowledge, this is the highest $Q$ demonstrated by MSW devices to date. The impedance at resonance is 237 $\Omega$, which translates to an impedance of 474 $\Omega$ for single YIG island resonator. The figure of merit (FOM) $f \times Q$ is $2.51 \times 10^{13}$ surpasses that of acoustic wave resonator counterparts [15].

From the measured 4.770-GHz resonance under 987-Oe magnetic bias, we could back-calculate that at 950 Oe, the resonance should be at 4.653 GHz using (1). The mismatch in resonant frequency is due to the fact that the effective magnetic bias from shape anisotropy and magnetocrystalline anisotropy was not accounted for in the simulation. The resonator can be modeled as a parallel $RLC$ circuit in series with a resistance and an inductance, as shown in Fig. 8. The series inductance and resistance are from the electrical loop inductance and the electrical resistance of the transducer, respectively. The parallel $RLC$ is the equivalent circuit for the resonance in the MSW domain. The extracted coupling factor is 0.206% $(k_t^2 = (\pi^2/4)((f_p - f_s)/f_p))$ with a resistance of 4.66 $\Omega$ from the inductor loop. This higher-than-expected electrical resistivity causes additional loading of the quality factor and explains the deviation of measured impedance (237 Ω) from the simulation result (810 Ω).

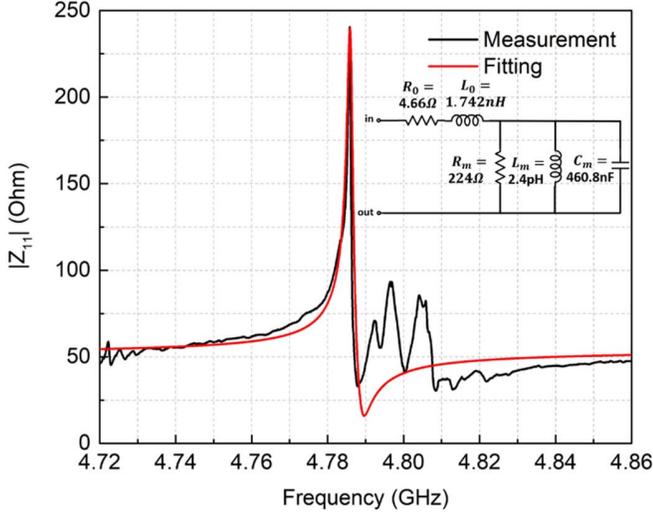

Fig. 8. Curve fitting of the resonant peak. Inset: equivalent circuit model.

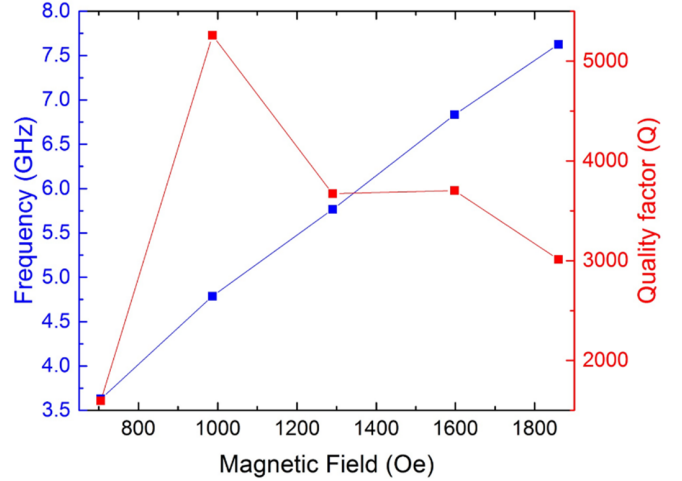

Fig. 10. MSW resonator resonance frequency and extracted quality factor.

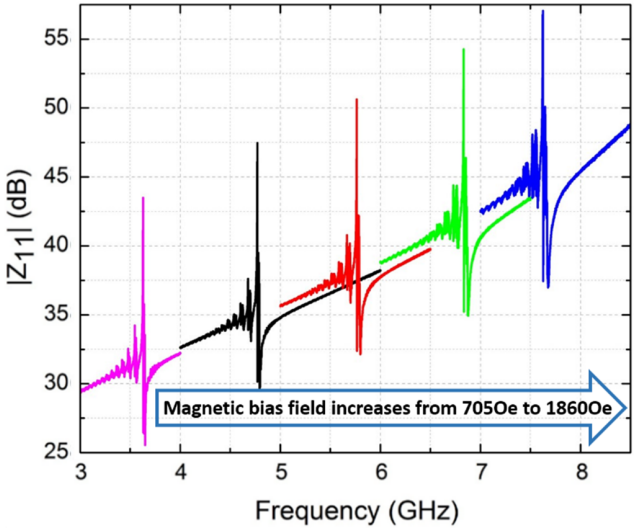

Fig. 9. Measured $|Z_{11}|$ of the MSW resonator under different magnetic bias fields from 987 to 1860 Oe.

$|Z_{11}|$ of the one-port MSW resonator under different magnetic biases is measured and plotted in Fig. 9. The dc magnetic bias was stepped from 705 to 1860 Oe, the frequency of the main resonance increased accordingly by more than one octave from 3.630 to 7.626 GHz. $Q$ calculated from the 3-dB bandwidth as a function of magnetic bias is plotted in Fig. 10, as well as the frequency of the main resonance a function of magnetic bias. As shown in Fig. 10, the tuning efficiency changes from 3.367 to 2.96 MHz/Oe as the magnetic bias increases from 705 to 1860 Oe, which is consistent with (2). The parasitic resonance from the pad's parasitic capacitance and the inductance of the loop generates a strong resonating current in the inductor, thus boosting the coupling of MSW. As the electrical impedance of MSW is determined by coupling and $Q$, this leads to large impedance variation, which is also reflected in Fig. 9. Interestingly, $Q$ varies with applied magnetic bias. At low bias, the device structure supports spin-wave modes that are half of the frequency of the main MBVW mode, and therefore, parametric spin-wave pumping by the MBVW is allowed, which degrades $Q$ [16]. As the bias increases, these half-frequency spin-wave modes are no longer permitted, and $Q$ increases. On the other hand, due to the unexpected high resistivity from the electrodes and the contact resistance, at even higher bias (therefore higher resonant frequency), the electrical resistance becomes significant and starts to limit $Q$. This can be prevented by switching to a different electrode material and better process control. Except in the spin-wave pumping regime, the extracted $Q$ valuers are consistently above 3000, and the highest $Q$ value achieved is 5259.

As predicted in simulations, a salient feature of MBVW resonators is that the spurious modes appear at the lower frequency side of the main resonance due to the abnormal dispersion of MBVW, which is quite similar to bulk acoustic wave devices with a type II dispersion [17].

The measured temperature coefficient of frequency (TCF) of the MSW resonator under different biases is shown in Fig. 11. Interestingly, the TCF varies from −996 to −1440 ppm/K as the dc magnetic bias decreases. This is because among the effects that contribute to TCF (such as temperature dependence of magnetocrystalline anisotropy, thermal stress-induced magnetocrystalline anisotropy, and magnetization), the dominating effect is the temperature dependence of the saturation magnetization. As shown previously, the resonant frequency can be approximated by $\omega = (\omega_0(\omega_0 + \omega_M))^{1/2}$ for $kd \ll 1$. Therefore, the TCF can be approximated by

$$\frac{d\omega}{dT} = \frac{1}{2} \frac{1}{1 + \frac{\omega_M}{\omega_0}} \cdot \frac{d\omega_M}{dT}. \qquad (3)$$

As shown in (3), the TCF increases as the dc bias decreases. Although the TCF of MSW devices is higher than that of typical BAW and SAW devices, which typically ranges from a few tens of ppm/K to ∼100 ppm/K, the temperature stability can be improved by leveraging the tunability of MSW devices with the tradeoff with system complexity. Table I summarizes the performance of the device presented in this article, and its comparison with state of the art. With our novel design and fabrication method, much higher $Q$ is achieved with much smaller sizes, and similar tuning efficiency and TCF.

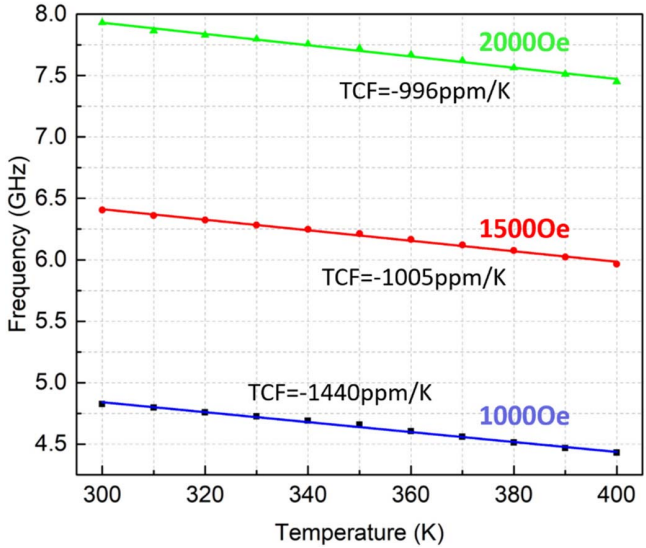

Fig. 11. Measured relationship between resonant frequency and temperature under different biases within the operating frequency range (with linear fitting).

TABLE I
COMPARISON OF THIS WORK WITH STATE-OF-THE-ART MSW RESONATORS

|  | This work | State-of-art |
|---|---|---|
| Size | 1.7mm*0.2mm | 25mm*3mm [19][20] |
| Tuning efficiency (MHz/Oe) | 2.96-3.37MHz/Oe | 2.8MHz/Oe for MFVW [18][19] 2.8MHz/Oe for MSSW [18] |
| Quality factor (Q) | >3,000@4.79-7.63GHz | 290-1570@2-11GHz [18] 500-530@2-4GHz [18] 375-1000@1-7GHz [19] ~800@2-4GHz [20] <600@2.5-3.5GHz [21] |
| Temperature coefficient of frequency (TCF)(ppm/K) | -996@4.7GHz -1005@6.2GHz -1440@7.7GHz | -1,375@3GHz for MFVW [18] +1140@8.05GHz for MSSW [18] |

## V. CONCLUSION

A novel MSW resonator structure consists of patterned YIG film and inductor loop transducer has been designed and fabricated with a novel microfabrication process on a chip. The lateral dimension of a single-patterned YIG structure is 1700 $\mu$m × 200 $\mu$m. The designed MSW resonator is tuned from 4.787 to 7.626 GHz, with measured quality factor at resonance frequency higher than 3000 across the whole tuning range. The $f \times Q$ product of these devices is significantly higher than their acoustic counterparts. The small device size made possible by our novel fabrication process and transducer design enables single-chip integration of multifrequency devices, which facilitates the realization of chip-scale high-order MSW filters, multiplexers, circulators [22], microwave-to-optical converters [23], and quantum coherent spin-magnon transducer [24]. The YIG micromachining process and resonator design are not limited to the GGG substrate. They can be directly ported to layer-transferred YIG thin-film technologies [25]. Leveraging the recent advances in integrated magnetic materials, a small permanent magnet (such as a screen-printed magnet [26]) is sufficient for providing homogeneous magnetic bias for these devices to operate in SHF bands because of the small size of these resonators, thus enabling a chip-scale SHF multiplexing solution.

ACKNOWLEDGMENT

The authors would like to thank the staff at Purdue's Birck Nanotechnology Center for their technical support. They would also like to thank Tingting Shen for discussions about fabrication and Yiyang Feng for discussions about SEM imaging.

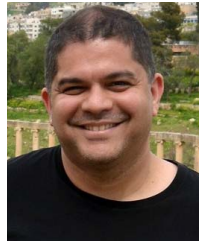

**Sunil A. Bhave** (Senior Member, IEEE) received the B.S. and Ph.D. degrees in electrical engineering and computer sciences from the University of California at Berkeley, Berkeley, CA, USA, in 1998 and 2004, respectively.

He was a Professor with Cornell University, Ithaca, NY, USA, for ten years and worked at Analog Devices, Woburn, MA, USA, for five years. In April 2015, he joined the Department of Electrical and Computer Engineering, Purdue University, West Lafayette, IN, USA, where he is currently the Associate Director of operations at the Birck Nanotechnology Center. He is a Co-Founder of Silicon Clocks, Fremont, CA, USA, that was acquired by Silicon Labs, Austin, TX, USA, in April 2010. His research interests focus on the interdomain coupling in optomechanical, spin-acoustic, and color center-MEMS devices.

Dr. Bhave received the NSF CAREER Award in 2007, the DARPA Young Faculty Award in 2008, the IEEE Ultrasonics Society's Young Investigator Award in 2014, and the Google Faculty Research Award in 2020. His students have received best paper awards at the IEEE Photonics 2012, the IEEE Ultrasonics Symposium 2009, and IEDM 2007.

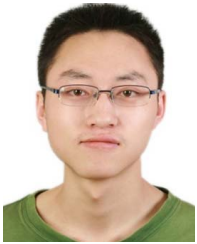

**Sen Dai** received the B.Sc. degree in physics from the University of Science and Technology of China, Hefei, Anhui, China, in 2014. He is currently pursuing the Ph.D. degree with the Department of Physics and Astronomy, Purdue University, West Lafayette, IN, USA.

He joined Prof. Sunil Bhave's OxideMEMS Lab in January 2017. His research is focused on RF MEMS resonators and micromachining ferrite components.

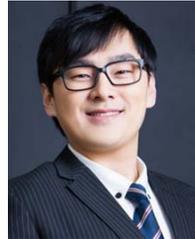

**Renyuan Wang** (Member, IEEE) received the B.S. degree from the Harbin Institute of Technology, Harbin, China, in 2007, the M.S. degree from the University of Massachusetts at Dartmouth, North Dartmouth, MA, USA, in 2010, and the Ph.D. degree from Cornell University, Ithaca, NY, USA, in 2014.

From 2015 to 2017, he was a Research and Development Engineer with the BAW RnD Group, Qorvo, Apopka, FL, USA. He then joined the FAST Labs, BAE Systems, Inc., Nashua, NH 03060 USA, where he is currently a Senior Scientist. He worked in developing high-dynamic-range coherent RF photonic radar front ends, lithium niobate thin-film devices for applications in RF MEMS, optomechanics, nonlinear optics, and inertial measurement units, as well as aluminum nitride bulk acoustic wave resonators and filters for personal mobile wireless devices. His current research interests focus on MEMS devices exploiting ferroelectric, ferro/ferrimagnetic materials, and their intercouplings.